\documentclass[
aps,
prc,
showpacs,
superscriptaddress,
nofootinbib,
floatfix]
{revtex4}
\usepackage{graphicx,
longtable}
\begin{document}
\title{Probing particle and nuclear physics models \\ of neutrinoless double beta decay with different
nuclei}
%
%
\author{        G.L.~Fogli}
\affiliation{   Dipartimento Interateneo di Fisica ``Michelangelo Merlin,'' 
               Via Amendola 173, 70126 Bari, Italy}
\affiliation{   Istituto Nazionale di Fisica Nucleare, Sezione di Bari, 
               Via Orabona 4, 70126 Bari, Italy}
\author{        E.~Lisi}
\affiliation{   Istituto Nazionale di Fisica Nucleare, Sezione di Bari, 
               Via Orabona 4, 70126 Bari, Italy}
\author{        A.M.~Rotunno}
\affiliation{   Dipartimento Interateneo di Fisica ``Michelangelo Merlin,'' 
               Via Amendola 173, 70126 Bari, Italy}
\affiliation{   Istituto Nazionale di Fisica Nucleare, Sezione di Bari, 
               Via Orabona 4, 70126 Bari, Italy}
\begin{abstract}
Half-life estimates for neutrinoless double beta  
decay depend on particle physics models for lepton-flavor violation, as well as 
on nuclear physics models for the structure and transitions of candidate nuclei. 
Different models considered in the literature can be
contrasted---via prospective data---with a ``standard'' scenario
characterized by light Majorana neutrino exchange and by the
quasiparticle random phase approximation, for which the theoretical
covariance matrix has been recently estimated. We show that, 
assuming future half-life data in four promising nuclei 
($^{76}$Ge, $^{82}$Se, $^{130}$Te, and $^{136}$Xe), 
the standard scenario can be distinguished from a few nonstandard physics models, 
while being compatible with alternative state-of-the-art nuclear calculations (at $95\%$ C.L.). 
Future signals in different nuclei may thus help to discriminate at least some decay mechanisms, 
without being spoiled by current nuclear uncertainties. 
Prospects for possible improvements are also discussed.
\end{abstract}
\medskip
\maketitle

\section{Introduction \label{SecI}}

The search for lepton number violation has gained new momentum after the
discovery of neutrino flavor transformations driven by $\nu$ masses and mixing \cite{PDGR}.
In particular, the search for neutrinoless double beta ($0\nu\beta\beta$) decay,
\begin{equation}
(Z,\, A)\to (Z+2,\, A) + 2e^- \ ,
\end{equation}
which violates the lepton number by two units, is at the forefront of current neutrino physics \cite{Av08}. 

For a given candidate nucleus $i=(Z,\,A)$, the $0\nu\beta\beta$ half-life $T_i$ can be expressed as 
\begin{equation}
\label{Ti}
T_i^{-1} = G^\ell_i\, |M_i^\ell|^2 \, \lambda_\ell^2\ ,
\end{equation}
where $G_i^\ell$ is an accurately calculable kinematical phase-space factor,  
$M_i^\ell$ is the $0\nu\beta\beta$ nuclear matrix element (NME), embedding
the nuclear physics aspects of the process, and $\lambda_\ell$ is a lepton number 
violation parameter, embedding the particle physics aspects of the process. The index $\ell$
labels different possible mechanisms underlying $0\nu\beta\beta$ decay \cite{Factor}.

If the three known neutrinos $\nu_i$ 
are described by Majorana (rather than Dirac) 4-spinors, then their masses $m_i$ 
and the $\nu_e$ mixing matrix elements $U_{ei}$ can generate one such mechanism, with $\lambda_\ell$
being equal to the so-called ``effective Majorana neutrino mass,'' $m_{\beta\beta}$ \cite{PDGR,Av08}  
\begin{equation}
\label{mbb}
m_{\beta\beta}=\left|\sum_{i=1}^3 m_i\,U^2_{ei}\right|\ .
\end{equation}
Apart from the above ``standard'' mechanism with the exchange of light Majorana neutrinos,  
$0\nu\beta\beta$ decays can also be generated by new particles and interactions (and even new
dimensions) beyond those of the standard electroweak model, as recently reviewed in \cite{DPas,Elli}.

Different models of particle and nuclear physics affect half-life predictions $T_i$ in
different ways, and may thus be discriminated, in principle, if 
the overall uncertainties are smaller than the spread of the predictions. In
particular, the use of half-life ratios,
\begin{equation}
\label{ratio}
\frac{T_i}{T_j}= \frac{G^\ell_j\, |M_j^\ell|^2}{G^\ell_i\, |M_i^\ell|^2 }\ , 
\end{equation}
has been advocated in order to effectively reduce the dependence on the following: ($i$) unknown particle physics parameters
\cite{Bile,Elli}, via cancellation of the factor $\lambda_\ell$; and ($ii$)  nuclear physics 
uncertainties \cite{DPas,Elli}, via cancellation of systematics in the NME ratio, so as to enhance 
the discrimination power of future data. In this context, the work \cite{Elli} represents (to our knowledge) 
the only quantitative study of particle and nuclear physics
model discrimination, based on prospective multi-isotope data and on guessed uncertainties---usually
dominated by theoretical NME errors.

Recently, there has been significant progress toward a reduction and a 
better understanding of the theoretical nuclear uncertainties. In particular, state-of-the-art 
NME calculations based on the quasiparticle random phase approximation (QRPA) \cite{Asse,Su08} or on the Shell Model
\cite{Shel} are increasingly converging within their estimated errors, 
at least in the most studied $0\nu\beta\beta$ case 
with light Majorana neutrino exchange. Moreover, within a specific QRPA model and its variants \cite{Asse}, the
theoretical error matrix has been estimated in detail \cite{Ours}, 
showing that high---and previously ignored---correlations
$\rho_{ij}$ exist between NME errors in any pair of nuclei $(i,\,j)$. 
High correlations imply that NME errors are largely cancelled in ratios $T_i/T_j$, thus motivating
a novel statistical analysis of prospective multi-isotope data where the $\rho_{ij}$ are explicitly
included---which is the goal of this work.

The structure of our paper is as follows. In Sec.~II we describe in detail our ``standard scenario,'' 
characterized by detailed QRPA estimates of NME's for light Majorana 
neutrino exchange \cite{Ours}. This scenario is used as null hypothesis in statistical tests, based
on prospective data for four $0\nu\beta\beta$ candidate nuclei 
($^{76}$Ge, $^{82}$Se, $^{130}$Te, and $^{136}$Xe). In Sec.~III we assume mock data according to
alternative particle physics model expectations, and show that two (out of seven) models can reject the
null hypothesis at $>95\%$ C.L. In Sec.~IV we repeat the exercise with different nuclear physics
models, and show that none of them can reject the null hypothesis at $95\%$ C.L. Therefore,
future $0\nu\beta\beta$ signals may indeed start to discriminate some underlying
particle physics mechanisms, without being spoiled by currently estimated nuclear physics
uncertainties. Theoretical error correlations are proven
to be crucial in this context.  Conclusions and prospects for future improvements are
discussed in Sec.~V.

\begin{table}[t]
\caption{Prospective half-life sensitivities at 90\% C.L. ($T_i^{90}$) for 
different nuclei $i$ in promising future projects, as reported in \protect\cite{Bara}.
All projects plan a second phase with lower backgrounds and higher sensitivies.}
\begin{ruledtabular}
\begin{tabular}{ccc}
$i$ & $T_i^{90}/\mathrm{y}$ & Project   \\[1mm]
\hline
$^{76}$Ge& $2.0\times10^{26}$  & GERDA, MAJORANA         \\
$^{82}$Se& $2.0\times10^{26}$  & SuperNEMO      \\
$^{130}$Te& $2.1\times10^{26}$  & CUORE   \\
$^{136}$Xe& $6.4\times10^{25}$  & EXO       
\end{tabular}
\end{ruledtabular}
\end{table}
\section{Null hypothesis and statistical approach}

In this work, we focus on four candidate nuclei for $0\nu\beta\beta$ decay, which will be probed in
promising, next-generation experiments \cite{Bara}: $^{76}$Ge, $^{82}$Se, $^{130}$Te, and $^{136}$Xe.
Table~I shows the expected half-life sensitivity in the first experimental phase, which might be followed by
an even more sensitive second phase \cite{Bara}. Among these four nuclei, $^{76}$Ge is perhaps the most studied,
both theoretically and experimentally. Therefore, its half-life $T_j$ 
may be used as an appropriate benchmark \cite{DPas,Elli}, namely,
as a common denominator in ratios $T_i/T_j$. For definiteness, we assume a prospective value
\begin{equation}
\label{norm}
T({}^{76}\mathrm{Ge})=10^{26}~\mathrm{y}\ ,
\end{equation}
reminding that such absolute half-life is immaterial and provides just a normalization. Our main results
do not depend on this specific choice for the ``denominator'' nucleus and its half-life. 
It is also appropriate  to linearize the analysis by taking logarithms of half-lives \cite{Ours}, 
\begin{equation}
\tau_i = \log_{10}(T_i/\mathrm{y})\ ,
\end{equation}
so as to deal with differences ($\tau_i-\tau_j$) rather than with ratios ($T_i/T_j$).

Recently, state-of-the-art calculations of $0\nu\beta\beta$ NME 
(within the QRPA and for light Majorana neutrino exchange \cite{Asse})  
have been thoroughly investigated from a statistical viewpoint \cite{Ours}, in order to assess their
inherent theoretical uncertainties. In particular, it has been shown that
the NME calculations display not only large variances (as is well known), but also large covariances among
any two nuclei, which can be accounted for by a correlation matrix $\rho_{ij}$. 
The errors and correlations estimated in \cite{Ours} originate from the following 
variants in the QRPA ingredients used as inputs:
($i$) two values for the axial 
coupling, namely, $g_A=1.25$ (bare) and $g_A=1.00$ (quenched); ($ii$) two approaches to short-range
correlations (s.r.c.), the so-called Jastrow-type s.r.c., and the unitary correlation operator method (UCOM); 
($iii$) three sizes for the model basis, small, 
intermediate, and large; ($iv$) two many-body models, namely, QRPA and
its renormalized version. All the 24 variants include 
the NME uncertainties induced by fitting the particle-particle strength
parameter $g_{pp}$ via the experimentally observed
$2\nu\beta\beta$ decay  process \cite{Asse}.

The final results of \cite{Ours}, together with the 
normalization in Eq.~(\ref{norm}), can be  translated easily into a set of
predictions for the logarithmic half-lives and their $\pm1\sigma$ errors, 
\begin{equation}
\tau^0_i \pm s^0_i\ (\mathrm{with\ correlations}\ \rho_{ij})\ , 
\end{equation}
where the superscript ($^0$) is intended to mark the ``null hypothesis'' in the following
statistical tests. Table~II shows the resulting
numerical values for our null hypothesis (or ``QRPA standard scenario'').

\begin{table}[t]
\caption{Null hypothesis: central values $\tau^0_i=\log_{10}(T_i/\mathrm{y})$, together with their
theoretical errors $s^0_i$ and correlations $\rho_{ij}$, in our standard QRPA scenario 
with light Majorana $\nu$ exchange 
\protect\cite{Ours}, normalized to a benchmark half-life $T_i/\mathrm{y}=10^{26}$  
for $i={}^{76}$Ge.}
\begin{ruledtabular}
\begin{tabular}{lrrrr}
Null hypothesis    & $^{76}$Ge  &$^{82}$Se  &$^{130}$Te  &$^{136}$Xe  \\[1mm]
\hline
QRPA, standard $\tau^0_i$ & 26.000 & 25.480 & 25.430 & 25.888   \\[1mm]
1-$\sigma$ error $s^0_i$     &  0.244 & 0.270  & 0.316  & 0.374    \\[1mm]
\hline
Correlations $\rho_{ij}$  & 1.000  & 0.978  & 0.899  & 0.805    \\[1mm]
                          &        & 1.000  & 0.927  & 0.846    \\[1mm]
                          &        &        & 1.000  & 0.916    \\[1mm]
                          &        &        &        & 1.000    \\
\end{tabular}
\end{ruledtabular}
\end{table}

The null hypothesis can be tested via prospective experimental data. We assume that several
experiments will observe positive $0\nu\beta\beta$ signals, and determine the decay lifetimes
with uncorrelated experimental errors, 
\begin{equation}
\tau_i \pm s_i\ (\mathrm{data})\ .
\end{equation}
Next-generation experiments might aim at a half-life accuracy of $\delta T_i/T_i \simeq  20\%$ \cite{Av08}, namely, 
\begin{equation}
\label{si}
 s_i \simeq 0.08\ ,
\end{equation}
which we adopt hereafter. The specific value of $s_i$ is not crucial, as far as
theoretical uncertainties remain dominant ($s^0_i\gg s_i$).
If the differences 
between standard predictions and mock data ($\tau^0_i-\tau_i$) 
are significantly larger than the theoretical errors ($s^0_i$)
in at least one nucleus (other than $^{76}$Ge, where $\tau^0_i=\tau_i=26.000$ by construction),
the null hypothesis will be rejected---or it will be accepted otherwise. 

The differences can be properly evaluated by $\chi^2$ statistics \cite{Ours},
\begin{equation}
\label{chi2}
\chi^2 = \sum_{i,j=1}^{N} (\tau_i-\tau^0_i)\,W_{ij}\,(\tau_j-\tau_j^0)\ ,
\end{equation}
where the inverse error matrix $W$ includes both theoretical errors (correlated) and
experimental errors (uncorrelated):
\begin{equation}
[W]^{-1}_{ij} = \rho_{ij}s^0_is^0_j+\delta_{ij}s_is_j\ . 
\end{equation}
In the above $\chi^2$ expression, $N$ is the number of nuclei considered in the analysis. 
We shall take $N=4$ by default, but also will show variations for $N=3$, whereas
the case with $N=2$ nuclei (providing a single half-life ratio with relatively
little constraining power) will not be commented in this work. 
Since the $^{76}$Ge half-life is used for normalization in all cases, the correct number of degrees of freedom 
is $N-1$. 
The rejection probability $P_r$ associated to the $\chi^2$ test \cite{Chi2} is thus
\begin{equation}
P_r=P_r(\chi^2;N-1)\ .
\end{equation}
Moreover, we set a threshold $P_r>95\%$ for a statistically 
significant rejection of the null hypothesis (implying, e.g., $\chi^2>7.8$ for $N=4$).

In comparison with the statistical approach of \cite{Elli}, our method uses a 
simple $\chi^2$ statistics (rather than Monte Carlo simulations), and includes 
theoretical error correlations (which will be proven to be crucial). On the other
hand, we test a single null hypothesis (the standard QRPA case,
where such correlations are well
defined \cite{Ours}), and focus on a smaller set of candidate nuclei than in \cite{Elli}.
In this sense, both our results and those in \cite{Elli} provide quantitative, but complementary, 
assessments of future  tests of  models for $0\nu\beta\beta$ decay.
Prospects for more comprehensive analyses, including the merits of both approaches, 
are discussed below in Sec.~V.

\begin{table}[t]
\caption{Estimated values of $\tau_i=\log_{10}(T_i/\mathrm{y})$ in various particle physics models for
$0\nu\beta\beta$ decay (different from light Majorana $\nu$ exchange), normalized to a  
benchmark half-life $T_i/\mathrm{y}=10^{26}$ for $i={}^{76}$Ge.}
\begin{ruledtabular}
\begin{tabular}{lccccc}
Particle physics model & $^{76}$Ge  &$^{82}$Se  &$^{130}$Te  &$^{136}$Xe  & Refs. \\[1mm]
\hline
Heavy~$\nu$       & 26.000 & 25.415 & 25.146 & 25.591 & \protect\cite{Heav} \\[1mm]
SUSY~$\pi$        & 26.000 & 25.431 & 25.462 & 25.863 & \protect\cite{SUPi} \\[1mm]
SUSY~${\tilde g}$ & 26.000 & 25.447 & 25.230 & 25.724 & \protect\cite{SUgl} \\[1mm]
RHC~$\eta$        & 26.000 & 25.462 & 25.301 & 25.732 & \protect\cite{RHCs} \\[1mm]
RHC~$\lambda$     & 26.000 & 25.146 & 25.114 & 25.826 & \protect\cite{RHCs} \\[1mm]
KK${+1}$          & 26.000 & 25.380 & 25.279 & 26.519 & \protect\cite{Kalu} \\[1mm]
KK${-1}$          & 26.000 & 25.415 & 25.255 & 25.892 & \protect\cite{Kalu} \\
\end{tabular}
\end{ruledtabular}
\end{table}
\begin{table}[ht]
\caption{Test of the null hypothesis (QRPA, standard), assuming 
$0\nu\beta\beta$  half-life data  as predicted by different particle physics models, with $20\%$ experimental uncertainties. Second column: rejection probability $P_r$ (\%) with four nuclei.
Third column: rejection probability range for three nuclei ($^{76}$Ge plus any other two). The null hypothesis 
can be rejected at $>95\%$ C.L.\ in two models, marked by *. Fourth and fifth columns: as in the previous two columns, but without correlations.}
\begin{ruledtabular}
\begin{tabular}{lrrrr}
Particle physics model &  4 nuclei, \%  & 3 nuclei, \% &  4 nuclei, \% & 3 nuclei, \%\\
& & & (no $\rho_{ij}$) & (no $\rho_{ij}$) \\[1mm]
\hline
Heavy~$\nu$       & 53.3 & 49.6 -- 71.3 & 29.8 & 28.0 -- 49.4 \\[1mm]
SUSY~$\pi$        &  5.2 &  5.6 -- 14.7 &  0.2 &  0.7 -- 2.0  \\[1mm]
SUSY~${\tilde g}$ & 27.1 & 18.8 -- 47.7 &  9.8 &  9.4 -- 24.4 \\[1mm]
RHC~$\eta$        & 10.5 & 17.9 -- 24.1 &  4.5 &  7.7 -- 14.9 \\[1mm]
RHC~$\lambda$ *   & 97.6 & 89.5 -- 97.9 & 50.1 & 38.3 -- 69.1 \\[1mm]
KK${+1}$      *   & 99.9 & 35.4 -- 99.9 & 61.8 & 15.7 -- 77.0 \\[1mm]
KK${-1}$          & 37.0 & 14.9 -- 57.2 &  4.8 &  2.6 -- 15.7 \\
\end{tabular}
\end{ruledtabular}
\end{table}

\section{Analysis of particle physics models}

Neutrinoless double beta decay can be triggered by several lepton-flavor mechanisms, involving
either long-range or short-range interactions. Although two or more mechanisms might be present
(and interfere) at the same time, we shall restrict ourselves to cases with just one dominant
contribution. In this context, several models have been recently reviewed in Refs.~\cite{DPas,Elli}, 
to which we refer the reader for details and references to earlier literature \cite{Moha}. 
We focus on the following nonstandard mechanisms for $0\nu \beta\beta $ decay: 
\begin{itemize}
\item[(1)] 		Exchange of Majorana neutrinos with heavy masses ($m_i>1$~GeV) \cite{Heav} (Heavy $\nu$);
\item[(2)] 	Supersymmetric models with R-parity violation and pion exchange \cite{SUPi} (SUSY~$\pi$);
\item[(3)] 	As above, but with gluino exchange \cite{SUgl} (SUSY~$\tilde g$);
\item[(4)] 	Left-right symmetric models with leptonic right-handed currents (RHC) coupled to 
                 hadronic left-handed  currents \cite{RHCs} (RHC~$\eta$); 
\item[(5)] 		As above, but coupled to hadronic right-handed currents \cite{RHCs} (RHC~$\lambda$); 
\item[(6)] 	Kaluza-Klein neutrino exchange via extra dimension with radius $R=(300~\mathrm{eV})^{-1}$ 
					and brane-shift parameter (in GeV$^{-1}$): $a=10^{+1}$ \cite{DPas,Kalu} (KK$+1$);
\item[(7)] 	As above, but with $a=10^{-1}$ \cite{DPas,Kalu} (KK$-1$). 
\end{itemize}
Another SUSY-accompanied mechanism listed in \cite{DPas} provides basically the same predictions as
the SUSY~$\tilde g$ model in our four reference nuclei, and thus it is not separately considered here.

Table~III shows the estimated logarithmic half-lives $\tau_i$ in the above particle physics models,
as derived from the numerical compilations in \cite{DPas,Elli}, but with the normalization in Eq.~(\ref{norm}). 
For each model, we assume 
the $\tau_i$'s as central values of mock data, with a putative experimental
uncertainty $s_i$ as in Eq.~(\ref{si}). We then test the null hypothesis in Table~II
via the $\chi^2$ test in Eq.~(\ref{chi2}). 

Before discussing quantitative results, we observe that the $\tau_i$ spread
induced by different models in Table~III is often comparable or smaller than the 
theoretical uncertainties $s^0_i$. However, there are also some outliers, e.g., a very high  
$^{136}$Xe half-life prediction  in the KK$+1$ model. Therefore, it should be possible 
to distinguish, at least, a few nonstandard physics models from the standard case
of light Majorana neutrino exchange (represented by
the null hypothesis).

Table~IV shows the results of our statistical analysis. Using all four nuclei, the rejection probability
for the null hypothesis 
ranges from an insignificant 5.2\% (SUSY~$\pi$ model)
to a highly significant 99.9\% (KK$+1$ model). Also another model (RHC~$\lambda$) provides a relatively high 
$P_r=97.6\%$. Therefore, with four nuclei, at least these two nonstandard $0\nu\beta\beta$ mechanisms
can be distinguished from the standard one at $>95\%$ C.L., while the other 
mechanisms are  statistically indistinguishable from the null hypothesis.

By discarding one nucleus (except the benchmark $^{76}$Ge one), 
the rejection level for $N-1=2$ degrees of freedom may change considerably in either directions,
depending on the role of that nucleus in the total $\chi^2$. In general, it may be advantageous 
to drop a nucleus whose standard and nonstandard half-life predictions are similar. Viceversa,
dropping a nucleus with rather different predictions may weaken the statistical power of the test. 

Table~IV also shows the  range of rejection probabilities $P_r$ with three nuclei.  
It turns out that the RHC~$\lambda$ and KK$+1$ models can reach $P>95\%$ not only with four nuclei,
but even with some three-nuclei combinations. In particular, we find that 95\% C.L.\ can be exceeded
by combining the benchmark Ge datum with: $(i)$ either Se+Te or Se+Xe data for the RHC-$\lambda$ model, and $(ii)$
either Xe+Se or Xe+Te data for the KK$+1$ model. In any case, 
with either three or four nuclei,
correlations of theoretical errors are crucial for the statistical power of the test: if they were neglected 
($\rho_{ij}=\delta_{ij}$), no model could be distinguished from the null hypothesis at a statistically
significant level (95\% or higher), as reported in Table~IV.

The role of theoretical error correlations can be better appreciated by means of Fig.~1, which
charts the logarithmic half-lives for all pairs of nuclei. The slanted ellipses represent the null hypothesis
(QRPA, standard), while the crosses represent the assumed mock data for the two most deviant models
(RHC~$\lambda$ and KK$+1$), within $1\sigma$ errors. By construction, all central values 
for $^{76}$Ge are aligned at $T=10^{26}$~y (vertical dotted lines). The relatively large ``distance''
between ``theory'' (ellipses) and ``data'' (crosses) depends crucially on the ellipse errors and
orientation. If correlations were dropped ($\rho_{ij}=\delta_{ij}$), all the ellipses would cover
a much larger fraction of the planes, and their frontier would get closer to the crosses, in general.
Viceversa, if correlations were maximal ($\rho_{ij}=1$), the ellipses would shrink to slanted
segments: theoretical uncertainties then would exactly cancel in half-life ratios, 
and the analysis would be dominated by experimental errors. 
This fact underlines the importance of detailed
estimates of the theoretical covariance matrix, in addition to other motivations
discussed in \cite{Ours}.

\begin{figure}[t]
\vspace*{+1.0cm}
\hspace*{0.cm}
\includegraphics[scale=.97]{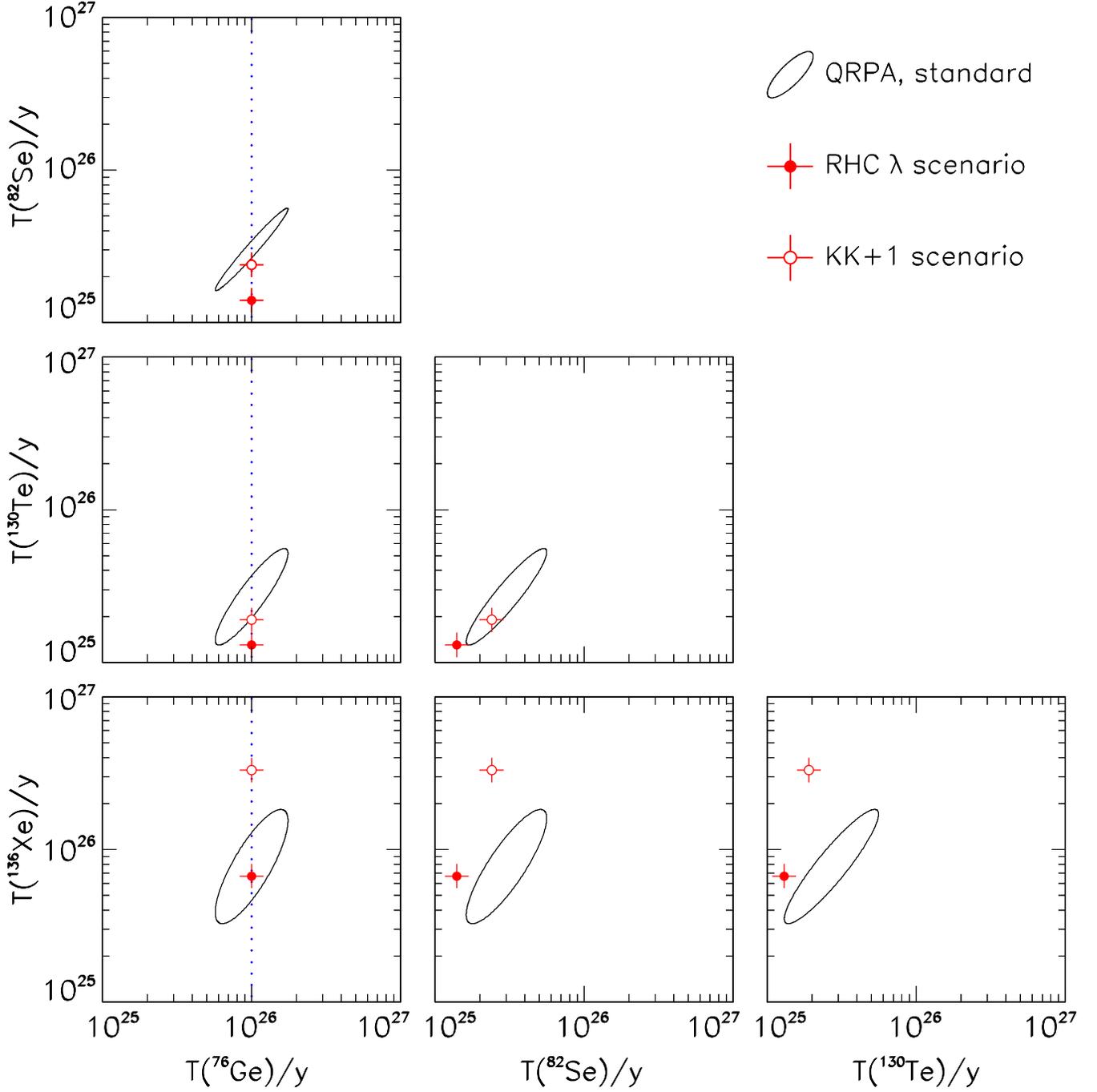}
\vspace*{+1.0cm}
\caption{ \label{f01} Projections of logarithmic 
half-life estimates in the coordinates planes of each pair of nuclei.
The ellipses represent our null hypothesis (QRPA, standard), with
evident correlations of theoretical errors (at $1\sigma$). 
The crosses represent mock data
(with assumed 20\% accuracy at $1\sigma$) 
centered on the predictions of two nonstandard
particle physics models described in the text, namely, RHC~$\lambda$ and 
KK$+1$. Data and predictions are conventionally aligned at a benchmark
value $T=10^{26}$~y for $^{76}$Ge (vertical dotted lines)}
\end{figure}

In conclusion, it appears that future measurements of $0\nu\beta\beta$ decay half-lives in
four (or even three) nuclei can discriminate the standard mechanism of light Majorana neutrino
exchange from at least the most deviant nonstandard mechanisms. The theoretical covariance
matrix plays a crucial role in the statistical analysis, and its control might improve the 
discrimination power of future measurements.

\section{Analysis of nuclear physics models}

The particle physics models considered in the previous Section (Tab.~III and IV) have
been implemented, in the available literature, within different approaches to the
nuclear physics structure and transitions for the $0\nu\beta\beta$ candidate nuclei considered.
Therefore, it makes sense also to analyze  the spread of predictions due to various nuclear
approximations. In this Section, we focus on the standard $0\nu\beta\beta$ physics mechanism 
(light Majorana $\nu$ exchange), and on a set of recent, 
state-of-the-art calculations of the associated nuclear matrix elements:
\begin{itemize}
\item[(1)] 	QRPA calculations by Suhonen and Kortelainen (S\&K) \cite{Su08} for fixed $g_A=1.00$ and
            Jastrow s.r.c.;
\item[(2)] 	QRPA (S\&K) for $g_A=1.00$ and UCOM s.r.c.;
\item[(3)] 	QRPA (S\&K) for $g_A=1.25$ and Jastrow s.r.c.;
\item[(4)] 	QRPA (S\&K) for $g_A=1.25$ and UCOM s.r.c.;
\item[(5)]  Shell Model calculations \cite{Shel}  for fixed $g_A=1.25$ and Jastrow s.r.c.;
\item[(6)] 	Shell Model with $g_A=1.25$ and UCOM s.r.c.  
\end{itemize}
See also the discussion in the Appendix of \cite{Ours}, where the NME from the above nuclear models
are compiled and compared with the NME used in our null hypothesis. We mention here that a third
approach, based on the microscopic interacting boson model (IBM) \cite{Iach}, provides NME's
in agreement with QRPA; however, the recent results in \cite{Iach} do not include $^{136}$Xe estimates
and thus are not included in this work.

Table~V shows the estimated logarithmic half-lives $\tau_i$ in the above six nuclear models,
as derived from the compilation in \cite{Ours} via the normalization in Eq.~(\ref{norm}). 
Analogously to the previous Section, for each model we assume 
the $\tau_i$'s as central values of mock data with errors $s_i$ [Eq.~(\ref{si})], and perform
a $\chi^2$  test of the null hypothesis of Table~II.

Table~VI shows the results of the statistical test, with either four or three nuclei, with and without theoretical
error correlations. It appears that, in any case,
none of the six nuclear models can reject the null hypothesis
(i.e., our reference QRPA calculations \cite{Asse}) at $\geq 95\%$ C.L. In other words, at such
level of significance, the 
six nuclear models considered are phenomenologically indistinguishable from our reference
QRPA model. This result is not in contradiction with the significant differences (up to $\sim 3\sigma$)
existing among the corresponding NME's \cite{Ours}. In fact, 
we are not really dealing with the absolute, unobservable NME, but only with their relative sizes 
(with respect to the benchmark $^{76}$Ge expectations), which exhibit a smaller spread.
Of course, the larger spread of absolute
NME will reappear elsewhere, namely, as a nuclear model
uncertainty in the lepton flavor violation parameter $\lambda_\ell$,
whose reconstruction uncertainties are beyond the scope of this work (see \cite{Ours} for a 
discussion).

\begin{table}[t]
\caption{Estimated values of $\tau_i=\log_{10}(T_i/\mathrm{y})$ in state-of-the-art nuclear physics models 
for $0\nu\beta\beta$ decay (for light Majorana $\nu$ exchange), 
 normalized to a benchmark half-life $T_i/\mathrm{y}=10^{26}$  for $i={}^{76}$Ge.}
\begin{ruledtabular}
\begin{tabular}{lccccc}
Nuclear physics model & $^{76}$Ge  &$^{82}$Se  &$^{130}$Te  &$^{136}$Xe  & Refs. \\[1mm]
\hline
QRPA (S\&K), $g_A=1.00$, Jastrow & 26.000 & 25.669 & 25.351 & 25.625 & \protect\cite{Su08} \\[1mm]
QRPA (S\&K), $g_A=1.00$, UCOM    & 26.000 & 25.663 & 25.319 & 25.621 & \protect\cite{Su08} \\[1mm]
QRPA (S\&K), $g_A=1.25$, Jastrow & 26.000 & 25.679 & 25.415 & 25.713 & \protect\cite{Su08} \\[1mm]
QRPA (S\&K), $g_A=1.25$, UCOM    & 26.000 & 25.669 & 25.363 & 25.689 & \protect\cite{Su08} \\[1mm]
Shell Model, $g_A=1.25$, Jastrow & 26.000 & 25.399 & 25.227 & 25.359 & \protect\cite{Shel} \\[1mm]
Shell Model, $g_A=1.25$, UCOM    & 26.000 & 25.407 & 25.207 & 25.343 & \protect\cite{Shel} \\
\end{tabular}
\end{ruledtabular}
\end{table}

\begin{table}[t]
\caption{Test of the null hypothesis (QRPA, standard), assuming 
$0\nu\beta\beta$  half-life data  as predicted by different nuclear physics models,  with $20\%$ experimental uncertainties. Second column: rejection probability $P_r$ (\%) with four nuclei.
Third column: rejection probability range for three nuclei ($^{76}$Ge plus any other two). The null hypothesis 
is not rejected, within $95\%$ C.L. Fourth and fifth columns: as in the previous two columns, but 
without correlations of theoretical errors.}
\begin{ruledtabular}
\begin{tabular}{lrrrr}
Nuclear physics model &  4 nuclei, \%  & 3 nuclei, \% &  4 nuclei, \% & 3 nuclei, \%\\
& & & $(\rho_{ij}=0)$ & $(\rho_{ij}=0)$ \\[1mm]
\hline
QRPA (S\&K), $g_A=1.00$, Jastrow & 86.8 & 45.0 -- 93.7 & 19.4 & 22.5 -- 37.0 \\[1mm]
QRPA (S\&K), $g_A=1.00$, UCOM    & 87.0 & 43.0 -- 93.1 & 20.5 & 23.6 -- 36.5 \\[1mm]
QRPA (S\&K), $g_A=1.25$, Jastrow & 79.9 & 30.8 -- 90.1 & 12.9 & 10.0 -- 29.8 \\[1mm]
QRPA (S\&K), $g_A=1.25$, UCOM    & 80.9 & 28.1 -- 89.9 & 14.2 & 14.5 -- 30.3 \\[1mm]
Shell Model, $g_A=1.25$, Jastrow & 78.9 & 46.4 -- 89.4 & 50.3 & 21.0 -- 68.4 \\[1mm]
Shell Model, $g_A=1.25$, UCOM    & 81.1 & 52.5 -- 90.7 & 53.6 & 23.5 -- 71.3 \\
\end{tabular}
\end{ruledtabular}
\end{table}

In Table~VI, the two independent models which deviate most from the null hypothesis are:
QRPA (S\&K) model with $g_A=1$ and Shell Model with $g_A=1.25$, both with UCOM s.r.c.
Figure~2 shows mock data according to such models (crosses), as compared with our standard
QRPA expectations (ellipses) in the same coordinate planes as in Fig.~1. 
The comparison of Figs.~1 and 2 shows at a glance that: $(i)$ the two most deviant nuclear physics model
(Fig.~2) are closer to standard predictions than the two most deviant particle physics models (Fig.~1);
and $(ii)$ in any give panel, the largest deviations between theory and data are generally opposite in 
Figs.~1 and 2, implying that nuclear physics variations do not mimic particle physics variations
in this context. 

We can thus conclude that, using prospective $0\nu\beta\beta$ data in four 
(and possibly just three) promising nuclei, the discrimination of some particle physics
mechanisms (in particular, RHC~$\lambda$ and KK$+1$ in
our analysis) is not spoiled by estimated uncertainties
in state-of-the-art nuclear theory calculations.
This is the main result of our work.

\begin{figure}[t]
\vspace*{+1.0cm}
\hspace*{0cm}
\includegraphics[scale=0.97]{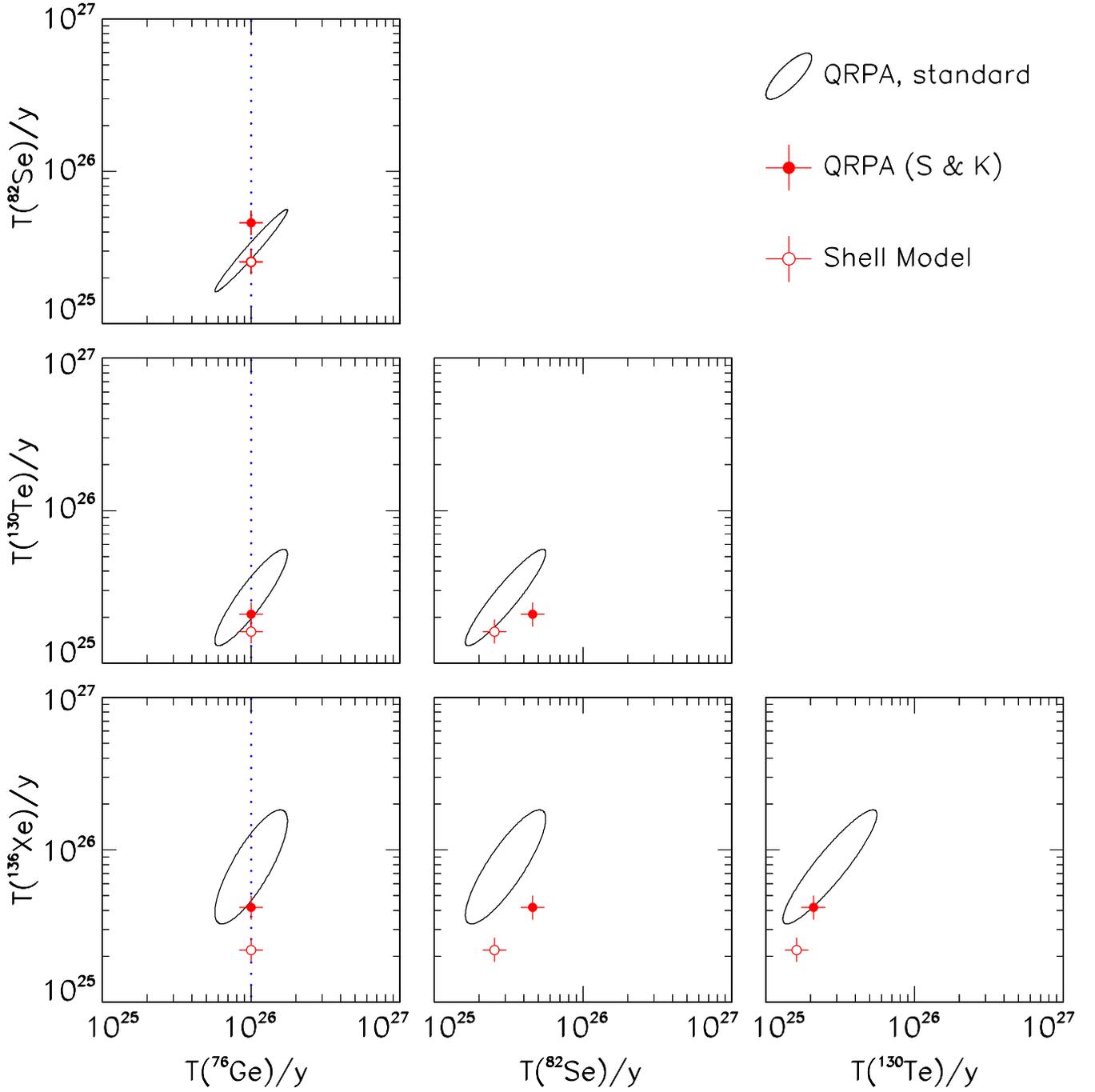}
\vspace*{+1.0cm}
\caption{ \label{f02} As in Fig.~1, but with mock data
centered on the predictions (for light Majorana $\nu$ exchange) 
of two alternative nuclear models, namely:
QRPA (S\&K) with $g_A=1$, and Shell Model with $g_A=1.25$, both with UCOM s.r.c.
See the text for details.}
\end{figure}
\newpage

\section{Summary and Discussion}

In this paper, we have focused on a reference $0\nu\beta\beta$ scenario
characterized by light Majorana neutrino exchange (concerning particle physics)
and by QRPA (concerning nuclear physics), for which the theoretical
covariance matrix can be defined in detail \cite{Ours}. We have used this scenario as 
a null hypothesis to be tested with prospective half-life data (with 20\% accuracy) 
in four promising nuclei 
($^{76}$Ge, $^{82}$Se, $^{130}$Te, and $^{136}$Xe), with the first nucleus being used as 
a conventional benchmark. Through our statistical approach, one
can properly quantify and implement the correlated uncertainties associated to half-life ratio tests
 of $0\nu\beta\beta$ decay,
which were proposed, e.g., in \cite{DPas,Elli} for particle physics models and in
\cite{Bile,Elli} for nuclear physics models. 

By setting a 95\% C.L.\ threshold for statistical significance, we have shown
that two nonstandard models for neutrinoless double beta decay, based
on right-handed leptonic currents and on Kaluza-Klein excitations in extra dimensions,
can reject the null hypothesis. Conversely, state-of-the-art nuclear physics models 
are consistent with the null hypothesis. Therefore, in case of positive 
$0\nu\beta\beta$ signals, next-generation  experiments will
have some discrimination power about the underlying particle physics mechanism,
despite known nuclear physics uncertainties. Since theoretical uncertainties 
dominate the analysis, their reduction can only increase the discrimination power of future
$0\nu\beta\beta$ data.  Correlations among theoretical
errors \cite{Ours} appear to be crucial ingredients in this context.

Our approach can be improved and extended in several ways, which we list in order
of increasing difficulty. Further prospective data can be easily included: 
the analysis does not need to be limited
to four nuclei (see, e.g., \cite{Elli}) and, in case of negative results, it might include 
one-sided limits besides two-sided signals. Analogously, further particle or nuclear physics
models can be included and tested. However,  
improvements on model discrimination power are less obvious, 
since theoretical covariances have been evaluated only for
a single set of QRPA calculations and for light Majorana neutrino exchange \cite{Ours}. 
For instance, in order to compare any two models as proposed 
in \cite{Elli} (rather than each model to the same null hypothesis), one should properly
attach a covariance matrix to each model. To reach this goal,
future NME estimates should include a ``statistical set'' of variants due to 
different input choices in their model parameter space, as also
emphasized in \cite{Ours}. It would also be useful to systematically
evaluate NME's for various particle physics mechanisms 
in one and the same nuclear physics model \cite{Priv} (and vice versa) in order to gauge
more precisely the impact of each model on observable half-lives, and to investigate
how many nuclei are needed for model discrimination at a given C.L. Needless to say, these goals
imply a long-term theoretical research program, which will be probably 
fully pursued only after unmistakable signals for $0\nu\beta\beta$ decay are found in two or more nuclei.
It is encouraging, however, to observe
that, at least for the standard case of light Majorana neutrino exchange, 
very different approaches (QRPA \cite{Asse,Inc1,Inc2}, Shell Model \cite{Shel} and IBM \cite{Iach})
are converging more than it could be hoped just a few years ago. Furthermore, within the
QRPA, the convergence improves as further nuclear structure constraints are included 
\cite{Inc1,Inc2}. First-order
electroweak reactions, probing the so-called first and second leg of $0\nu\beta\beta$ decay, 
might also help to reduce parametric uncertainties of
nuclear models  \cite{Ours,Zube}.

Finally, when $0\nu\beta\beta$ half-life data will be hopefully established in multiple nuclei,
additional tests of the underlying particle physics mechanism might include further 
constraints from the kinematical distributions of the emitted electrons \cite{Do83,Al06}, from the
comparison of $0\nu\beta\beta$ decay with
electron capture \cite{Hi94}, from branching ratios of $0\nu\beta\beta$ decays to ground and excited states
\cite{Exci},
and from possible links with other lepton-flavor violating processes (e.g., $\mu\to e)$ \cite{Ciri}. 
Such a rich phenomenology will then warrant more comprehensive analyses than 
the one proposed in this work.

\acknowledgments

This work is supported in part the Italian Istituto Nazionale di Fisica 
Nucleare (INFN) and Ministero dell'Istruzione, dell'Universit\`a e della Ricerca 
(MIUR) through the ``Astroparticle Physics'' 
research project. It is also supported in part by the EU ILIAS project through the
European Network of Theoretical Astroparticle Physics (ENTApP). 
We thank F.~Simkovic for interesting discussions about tests of nonstandard $0\nu\beta\beta$
mechanisms, which motivated us to undertake this work.



\begin{thebibliography}{99}

\bibitem{PDGR}  C.~Amsler {\it et al.}  [Particle Data Group],
 				``Review of Particle Physics,''
 				Phys.\ Lett.\  B {\bf 667}, 1 (2008).


\bibitem{Av08}  F.~T.~Avignone~III, S.~R.~Elliott and J.~Engel,
    			``Double Beta Decay, Majorana Neutrinos, and Neutrino Mass,''
               Rev.\ Mod.\ Phys.\  {\bf 80}, 481 (2008)
               [arXiv:0708.1033 [nucl-ex]].
               

 
\bibitem{Factor} The factorization of $T_i$ implicit in Eq.~(\protect\ref{Ti}) may not hold for
mechanisms involving particle masses of the same order
of the characteristic nuclear scale ($\sim\!100$~MeV) of the process. See, e.g., 
the active-sterile neutrino mechanism described in: P.\ Bamert, C.P.\ Burgess, and R.N.\ Mohapatra,
Nucl.\ Phys.\ B {\bf 438}, 3 (1995); P.\ Benes, A.\ Faessler, S.\ Kovalenko, and F.\ Simkovic,
Phys.\ Rev.\ D {\bf 71}, 077901 (2005). In any case, the numerical results of this work are
directly based on $T_i$ estimates (not on their factorization) in each nucleus.
 


\bibitem{DPas}  F.~Deppisch and H.~Pas,
  ``Pinning down the mechanism of neutrinoless double beta decay with
  measurements in different nuclei,''
  Phys.\ Rev.\ Lett.\  {\bf 98}, 232501 (2007)
  [arXiv:hep-ph/0612165].

\bibitem{Elli}  V.~M.~Gehman and S.~R.~Elliott,
  ``Multiple-isotope comparison for determining $0\nu \beta \beta$ decay mechanisms,''
  J.\ Phys.\ G {\bf 34}, 667 (2007)
  [Erratum-ibid.\  {\bf G35}, 029701 (2008)]
  [arXiv:hep-ph/0701099].

\bibitem{Bile}  S.~M.~Bilenky and J.~A.~Grifols,
 				``The possible test of the calculations of nuclear matrix elements of the
 				$0\nu\beta\beta$ decay,''
 				Phys.\ Lett.\  B {\bf 550}, 154 (2002)
 				[arXiv:hep-ph/0211101]; 
				  S.~M.~Bilenky and S.~T.~Petcov,
 				``Nuclear matrix elements of $0\nu\beta\beta$-decay: Possible test of the calculations,''
 				arXiv:hep-ph/0405237. The usefulness of multi-isotope data was also mentioned in:
				  S.~Pascoli, S.~T.~Petcov and L.~Wolfenstein,
                ``Searching for the CP-violation associated with Majorana neutrinos,''
                Phys.\ Lett.\  B {\bf 524}, 319 (2002)
                [arXiv:hep-ph/0110287].

\bibitem{Asse}   V.~A.~Rodin, A.~Faessler, F.~Simkovic and P.~Vogel,
                ``Assessment of uncertainties in QRPA $0\nu\beta\beta$-decay nuclear matrix
                elements,''
                Nucl.\ Phys.\  A {\bf 766}, 107 (2006)
                [Erratum-ibid.\  A {\bf 793}, 213 (2007)]
                [arXiv:0706.4304 [nucl-th]];
                F.~Simkovic, A.~Faessler, V.~Rodin, P.~Vogel and J.~Engel,
 				``Anatomy of nuclear matrix elements for neutrinoless double-beta decay,''
 				Phys.\ Rev.\  C {\bf 77}, 045503 (2008)
 				[arXiv:0710.2055 [nucl-th]].

\bibitem{Su08}  J.~Suhonen and M.~Kortelainen,
 				``Nuclear matrix elements for double beta decay,''
 				Int.\ J.\ Mod.\ Phys.\  E {\bf 17}, 1 (2008).


\bibitem{Shel}	J.~Menendez, A.~Poves, E.~Caurier and F.~Nowacki,
		``Disassembling the Nuclear Matrix Elements of the Neutrinoless double beta Decay,''
  			Nucl.\ Phys.\  A {\bf 818}, 139 (2009)
  			[arXiv:0801.3760 [nucl-th]]; 				
			``Deformation and the Nuclear Matrix Elements of the Neutrinoless Double Beta Decay,''
 				arXiv:0809.2183 [nucl-th].


\bibitem{Ours} A.~Faessler, G.~L.~Fogli, E.~Lisi, V.~Rodin, A.~M.~Rotunno and F.~Simkovic,
 ``Quasiparticle random phase approximation uncertainties and their correlations in the analysis of neutrinoless double beta decay,'' Phys.\ Rev.\ D {\bf 79}, 053001 (2009)
    [arXiv:0810.5733 [hep-ph]].



\bibitem{Bara}  A.~S.~Barabash,
 				``Double beta decay: present status,''
  				Proceedings of the 13th Lomonosov Conference
   				on Elementary Particle Physics (Moscow, Russia, 2007),
               ed.\ by A.I.\ Studenikin (World Scientific, Singapore, 2009)				
               [arXiv:0807.2948 [hep-ex]].


\bibitem{Chi2}  See, e.g., the CERN Library routine {\tt PROB} (G100),
                website: cernlib.web.cern.ch


\bibitem{Moha}
For the sake of brevity, we refer only to relatively recent literature.
However, some nonstandard mechanisms were introduced long ago; see e.g.:
R.~N.~Mohapatra, ``New contributions to neutrinoless double-beta decay in supersymmetric
theories,'' Phys.\ Rev.\  D {\bf 34}, 3457 (1986);
R.~N.~Mohapatra,
``Limits on the mass of the right-handed Majorana neutrino,''
 Phys.\ Rev.\  D {\bf 34}, 909 (1986). 
 
 
\bibitem{Heav}    F.~Simkovic, G.~Pantis, J.~D.~Vergados and A.~Faessler,
  ``Additional Nucleon Current Contributions to Neutrinoless Double Beta Decay,''
  Phys.\ Rev.\  C {\bf 60}, 055502 (1999)
  [arXiv:hep-ph/9905509].

\bibitem{SUPi}    A.~Faessler, S.~Kovalenko and F.~Simkovic,
  ``Pions in nuclei and manifestations of supersymmetry in neutrinoless  double beta decay,''
  Phys.\ Rev.\  D {\bf 58}, 115004 (1998)
  [arXiv:hep-ph/9803253].


\bibitem{SUgl}   M.~Hirsch, H.~V.~Klapdor-Kleingrothaus and S.~G.~Kovalenko,
  ``Supersymmetry and neutrinoless double beta decay,''
  Phys.\ Rev.\  D {\bf 53}, 1329 (1996)
  [arXiv:hep-ph/9502385].

\bibitem{RHCs}   K.~Muto, E.~Bender and H.~V.~Klapdor,
  ``Nuclear structure effects on the neutrinoless double beta decay,''
  Z.\ Phys.\  A {\bf 334}, 187 (1989).


\bibitem{Kalu}  G.~Bhattacharyya, H.~V.~Klapdor-Kleingrothaus, H.~Pas and A.~Pilaftsis,
  ``Neutrinoless double beta decay from singlet neutrinos in extra
  dimensions,''
  Phys.\ Rev.\  D {\bf 67}, 113001 (2003)
  [arXiv:hep-ph/0212169].


\bibitem{Iach} J.\ Barea and F.\ Iachello,
  ``Neutrinoless double beta decay in the microscopic interacting boson model,''
	Phys.\ Rev.\ C {\bf 79}, 044301 (2009).

\bibitem{Priv} F.~Simkovic, private communication.


\bibitem{Inc1}
  J.~Suhonen and O.~Civitarese,
  ``Effects of orbital occupancies on the neutrinoless beta beta matrix element
  of Ge-76,''
  Phys.\ Lett.\  B {\bf 668}, 277 (2008).

\bibitem{Inc2}
  F.~Simkovic, A.~Faessler and P.~Vogel,
  ``$0\nu\beta\beta$ nuclear matrix elements and the occupancy of individual
  orbits,''
  Phys.\ Rev.\  C {\bf 79}, 015502 (2009)
  [arXiv:0812.0348 [nucl-th]].
  
\bibitem{Zube}	  K.~Zuber,
 				``Summary of the workshop on 'Matrix elements for neutrinoless double beta
 				decay','' presented at
				IPPP Workshop on Matrix Elements for Neutrinoless Double Beta Decay (Durham, England, 2005). 
 				arXiv:nucl-ex/0511009.


\bibitem{Do83} M.~Doi, T.~Kotani, H.~Nishiura and E.~Takasugi,
  ``The Energy Spectrum And The Angular Correlation In The Beta Beta Decay,''
  Prog.\ Theor.\ Phys.\  {\bf 70}, 1353 (1983).

\bibitem{Al06}   A.~Ali, A.~V.~Borisov and D.~V.~Zhuridov,
  ``Probing new physics in the Neutrinoless double beta decay using electron
  angular correlation,''
  Phys.\ Rev.\  D {\bf 76}, 093009 (2007)
  [arXiv:0706.4165 [hep-ph]].

\bibitem{Hi94}
 M.~Hirsch, K.~Muto, T.~Oda and H.~V.~Klapdor-Kleingrothaus,
  ``Nuclear structure calculations of $\beta^+\beta^+$, $\beta^+$/EC and EC/EC
  decay matrix elements,''
  Z.\ Phys.\  A {\bf 347}, 151 (1994).

\bibitem{Exci} F.~Simkovic, M.~Nowak, W.~A.~Kaminski, A.~A.~Raduta and A.~Faessler,
  ``Neutrinoless double beta decay of Ge-76, Se-82, Mo-100 and Xe-136 to
  excited 0+ states,''
  Phys.\ Rev.\  C {\bf 64}, 035501 (2001)
  [arXiv:nucl-th/0107016].
  
\bibitem{Ciri}  
    V.~Cirigliano, A.~Kurylov, M.~J.~Ramsey-Musolf and P.~Vogel,
  ``Neutrinoless double-beta decay and lepton flavor violation,''
  Phys.\ Rev.\ Lett.\  {\bf 93}, 231802 (2004)
  [arXiv:hep-ph/0406199].

\end{thebibliography}
\end{document}